\documentclass[twocolumn,9pt,oneside,a4paper,DIV15]{scrartcl}
\usepackage[pdftex]{graphicx} \pdfcompresslevel=9
\usepackage[utf8]{inputenc}
\usepackage[T1]{fontenc}
\usepackage{cite}
\usepackage{fancyhdr}
\usepackage{abstract}

\author{Claudius~Jähn}
\title{Progressive Blue Surfels}

\begin{document}
\twocolumn[
\maketitle
\begin{onecolabstract}
In this paper we describe a new technique to generate and use surfels for rendering of highly complex, polygonal 3D scenes in real time. 
The basic idea is to approximate complex parts of the scene by rendering a set of points (surfels). 
The points are computed in a preprocessing step and offer two important properties: 
They are placed only on the visible surface of the scene's geometry and they are distributed and sorted in such a way, that every prefix of points is a good visual representation of the approximated part of the scene. 
An early evaluation of the method shows that it is capable of rendering scenes consisting of several billions of triangles with high image quality. \\
\emph{Please note that this paper represents a working draft, which will be subsequently replaced by refined versions!}
\end{onecolabstract}
]

\section{Introduction}
One way of rendering complex polygonal 3d scenes is to replace parts of the scene's geometry by a set of points.
This idea is for example used by Rusinkiewicz for the QSplat-technique \cite{Rusinkiewicz2000} for rendering large meshes.
Pfister et al. \cite{Pfister2000} introduce surfels for rendering complex objects (including texturing). 
Coconu and Hege \cite{Coconu2002} further try to exploit the capabilities of the graphics hardware to support rendering huge scene consisting of many complex objects.

\begin{figure*}[htbp]
	\centering
		\includegraphics[width=1.00\textwidth]{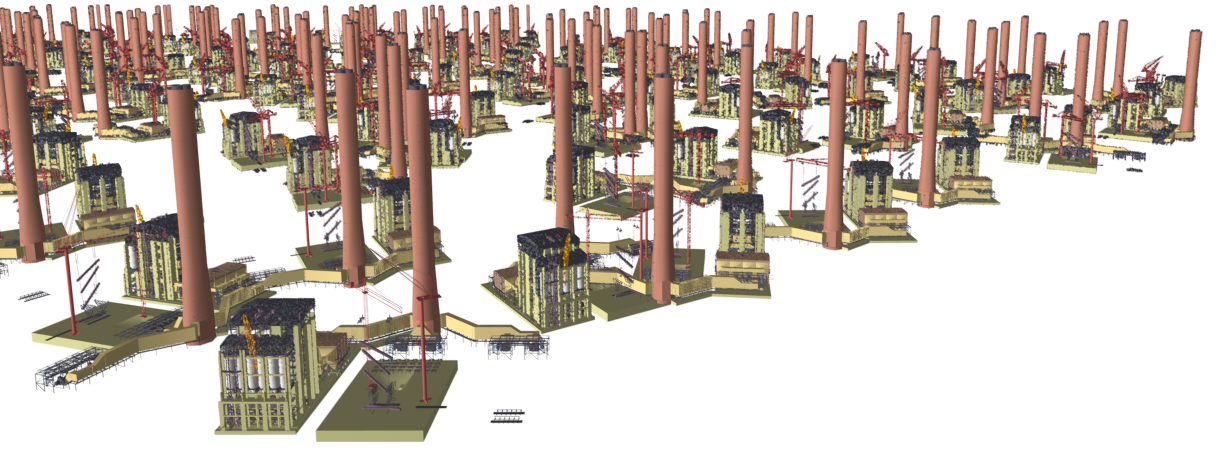}
	\caption{Example scene consisting of several Power Plant models}
	\label{fig:example}
\end{figure*}

In this paper we present the \emph{Progressive Blue Surfels} technique, which based on similar ideas, but has the following feature:
\begin{itemize}
	\item Support for huge hierarchical scenes (consisting of billions of polygons, see Figure \ref{fig:example} for an example)
	\item Objects inside the hierarchy can be instances
	\item Texture support
	\item Almost no popping artifacts even without alpha blending
	\item Simple to implement
\end{itemize}

\begin{figure*}[htbp]
	\centering
		\includegraphics[width=1.00\textwidth]{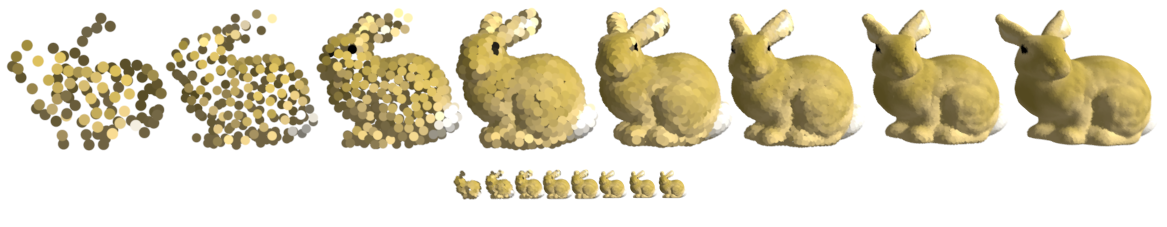}
	\caption{(top) Rendering of prefixes of different lengths (first 100, 200, 400, 800, 1.6~k, 3.2~k, 6.4~k, 12.8~k surfels). Note that the Point size is reduced for higher number of points.
		 (bottom) Rendering of the same prefixes from a larger distance.}
	\label{fig:prefixes}
\end{figure*}

The main idea of the method is:
Create an array of surfels for each node in the scene graph for which the complexity exceeds some reasonable value.
The surfels in an array are ordered in such a way, that each prefix of the array is a good graphical approximation of the subtree (see Figure \ref{fig:prefixes}).
The goal is, that the minimal distance between any two points is maximized for each prefix (targeting a blue noise distribution of the points).
This is done using an adaptive sampling approach (see Section \ref{sec:sampling}).
At runtime, the scene graph is traversed beginning at the root node (see Section \ref{sec:rendering}). 
For nodes having a small projected size (and an associated array of surfels), a prefix of the surfel array is rendered to approximate the subtree's geometry.
The number of surfels is chosen relatively to the node's projected size.
Other inner nodes are traversed further and other (large or simple) leaf nodes are rendered using their original geometry.

%

\section{Preprocessing: Generating the Surfels}
The input of the preprocessing step is the virtual scene, organized in a hierarchically organized scene graph and few parameters (described later).
If the scene itself does not define such a structure, a loose octree can be used to structure the scene's objects (or triangles) into a suitable structure.
Each node in the scene graph is evaluated.
If a node's subtree has a high complexity (more than e.g. 10000 polyogons) surfels are created as described in the next sections.
The surfels for a node are stored as one primitive object (e.g. a vertex buffer object (VBO) in OpenGL) and attached to the node as an attribute.

\subsection{Create the initial Set of Surfels}
\begin{figure*}[htbp]
	\centering
		\includegraphics[width=1.00\columnwidth]{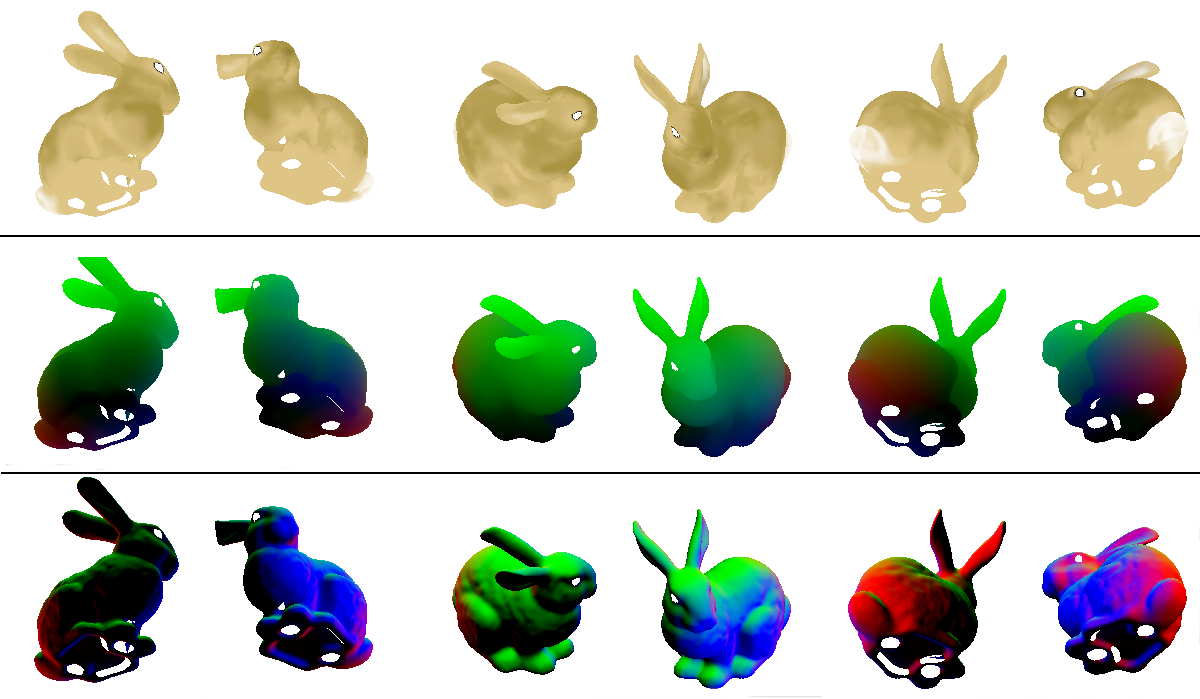}
	\caption{Content of the diffuse, position, and normal buffer during the surfel creation of the bunny model}
	\label{fig:rasterization}
\end{figure*}

The first step to create the surfels for a node is to create the \emph{initial set of surfels}, that is later reduced and ordered (see Section \ref{sec:sampling}).
It consist of points described by a positions normal, and a material description (as required for the rendering pipeline: e.g. a single color or ambient, diffuse, specular colors).
The points are created by rendering the selected node's subtree into multiple buffers -- using a shader to encode the different properties (color, position, normal, etc.) into the color of one buffer each.
For this, we take eight camera positions (the corners of the bounding box enclosing the subtree, looking at the center) under orthographic projection (see Figure \ref{fig:rasterization}). 
The scaling is chosen so that the larger side of the projected bounding box fits a chosen \emph{resolution parameter}.
(Note: This resolution parameter should be chosen larger than the projected side length for which surfels should be rendered at runtime.)
All occupied pixels are extracted from the buffers and combined into the initial set of surfels (a surfel is a tupel of property values).
Additionally, an estimator value for the relative covered area of the node is calculated -- the number of occupied pixels divided by the projected size of the used bounding boxes.


\subsection{Select and sort Surfels by Dart Throwing}
\label{sec:sampling}
In this step, the initial set of surfels is filtered and ordered to represent a progressive LOD of the represented node.
Beside the set of surfels, a value for the \emph{maximum number of surfels} is required as parameter.

The selection algorithm works as follows:
The initial surfels are stored in an input set.
For the ordered surfels, an output array is reserved -- initially containing one randomly chosen surfel from the input set.
Until the output array contains the maximal number of surfels or the input set is empty, the following steps are repeated:
A constant number (e.g. 200) of random surfels from the input set are chosen as candidates (Note: this is a simplification and must eventually be described in more detail!).
For all candidates, the euclidean distance to the closest surfel in the output array is computed.
 (This can be done efficiently by internally using an octree for storing the positions of the surfels in the output array.)
The candidate with the largest distance is extracted from the input set and appended to the output array.
The candidate with the smallest distance is extracted and discarded. 
The other candidates remain in the input set.
Then the process restarts.

The result of this sampling algorithm is, that the closest pairwise distances of the chosen points gradually decrease (on average, not strictly) while surfels lying too close to an already chosen surfel are filtered out.
The latter speeds up the overall sampling process: 
Without this filtering, the number of required candidates had to increase during the sampling process in order to still acquire relevant candidates.

\section{Rendering}
\label{sec:rendering}
The rough overview of the rendering process at runtime is as follows:

The scene graph is traversed from top to bottom.
If the projected size of a node's bounding box is above a given threshold or the node is not attached with an array of surfels, the traversal is continued (or the node is rendered if  it contains geometry).
For small nodes, a prefix of the surfels are rendered using point rendering.
The size of the prefix is chosen by the projected size of the node's bounding box in pixels, multiplied by the estimator value for the node's relative covered area (calculated in the preprocessing phase), and multiplied by an overdraw factor.
The overdraw factor must be chosen high enough to prevent the occurrence of holes (a value of 4 is typical).

To prevent popping artifacts when switching between original geometry and points or between points in different levels, both are rendered in a transition phase.
The size of the prefix of the one level decreases after the maximal prefix of the next level is rendered.

The size of the points can be increased to several pixels in order to allow rendering with fewer points without creating holes.
It is then advisable to use a vertex shader for rendering the surfels that adjusts the individuals surfel's size by the direction of the eye space normal (surfels having an orthogonal normal should have a smaller size).
Otherwise, the node's silhouette can severely be disturbed.

\section{Experimental Evaluation}
The most important points for an evaluation of the presented technique are:
\begin{itemize}
	\item the analysis of the blue noise property achieved by the sampling approach,
	\item the frame rate in dependency to the scene's complexity,
	\item and the achieved image quality compared to non approximate rendering.
\end{itemize}
The experimental evaluation is in progress and will eventually be added to this document.


\section{Conclusion}
We presented a point based rendering technique for complex polygonal scenes.

One possibility to improve the distribution of the sample points on a wider range of surfaces is to maximize the geodesic distance of the sample points instead of the euclidean distance (like done by Bowers et al. \cite{Bowers2010}).

Besides the rendering of scenes on a single workstation, the principle of the ordered surfels makes it a good candidate for streaming applications.
While the surfel arrays are transmitted, the already received prefix can by used as an approximation for the corresponding scene part. 
Further transmissions then refine the approximation progressively.

\bibliographystyle{alpha} 
\bibliography{../Articles/database/Articles}

\end{document}